\documentclass{PoS}
\title{Dense Quark Matter in a Magnetic Field}
\author{Efrain J. Ferrer,  \speaker{Vivian de la Incera}\\
Western Illinois University, USA \\
E-mail: \email{ej-ferrer@wiu.edu}, \email{v-incera@wiu.edu}}
\author{ Cristina Manuel\\
Instituto de Fisica Corpuscular (CSIC-U. de Valencia), Spain\\
E-mail: \email{Cristina.Manuel@ific.uv.es}} \abstract{We explore the
effects of an applied strong external magnetic field in the
structure and magnitude of the color superconducting diquark
condensate of a three massless flavor theory . The long-range
component of the B field that penetrates the superconductor enhances
the condensates formed by quarks charged with respect to this
electromagnetic field.} \FullConference{29th Johns Hopkins Workshop
in Theoretical Physics: Strong Matter in the Heavens. Budapest,
August 1-3, 2005} \ShortTitle{Dense Quark Matter in a Magnetic
Field}
\begin{document}
\section{Introduction}
It is well established that at high baryon density the combination
of asymptotic freedom and the existence of attractive channels in
the color interaction between the quarks lying in the large Fermi
surface come together to promote the formation of quark-quark pairs,
which in turn break the color gauge symmetry giving rise to the
phenomenon of color superconductivity. At densities much higher than
the masses of the u, d, and s quarks, one can assume the three
quarks as massless and the favored state results to be the so-called
Color-Flavor-Locking (CFL) phase \cite{alf-raj-wil-99/537},
characterized by a spin zero diquark condensate antisymmetric in
both color and flavor.

The conditions of extremely large density and very low temperature
required for color superconductivity cannot be recreated in Earth's
labs. Fortunately, nature provides us with a laboratory to probe
color superconductivity, the cores of celestial compact objects.
These compact stars typically have very large magnetic fields.
Neutron stars can have magnetic fields as large as $B \sim 10^{12} -
10^{14}$ G in their surfaces, while in magnetars they are in the
range $B \sim 10^{14} - 10^{15}$ G, and perhaps as high as $10^{16}$
G \cite{magnetars} (for a recent review of magnetic fields in dense
stars see \cite{lugones/0504454}). Even though we do not know yet of
any suitable mechanism to produce more intense fields, the virial
theorem \cite{virial} allows the field magnitude to reach values as
large as $10^{18}-10^{19}$ G. If quark stars are self-bound rather
than gravitational-bound objects, the upper limit that has been
obtained by comparing the magnetic and gravitational energies, could
go even higher.

A natural question to ask is: What is the effect, if any, of the
huge star's magnetic field in the color superconducting core? A
complete answer to this question would require a rather involved
study of quark matter at the intermediate range of densities proper
of neutron stars, where the strange quark mass cannot be ignored,
with the additional complication of an extra parameter, the magnetic
field. However, as a first, more tractable approach to this
question, one can ignore the strange quark mass effects and look for
the consequences of an external magnetic field on the
superconducting phase, assuming that the quark matter is formed by
three massless flavors. This was the strategy followed in our recent
paper \cite{MCFL}, whose main results will be described in what
follows.

In this talk I will show the way a magnetic field affects the
pairing structure and hence its symmetry, ultimately producing a
different superconducting phase that we have called Magnetic
Color-Flavor-Locking (MCFL) phase.

In a conventional superconductor, since Cooper pairs are
electrically charged, the electromagnetic gauge invariance is
spontaneously broken, thus the photon acquires a Meissner mass that
can screen a weak magnetic field, the phenomenon of Meissner effect.
In spin-zero color superconductivity, although the color condensate
has non-zero electric charge, there is a linear combination of the
photon and a gluon that remains massless \cite{alf-raj-wil-99/537}.
This new field plays the role of the "in-medium" photon in the color
superconductor, so the propagation of light in the color
superconductor is different from that in an electric superconductor.

Because of the long-range "rotated" electromagnetic field, a
spin-zero color superconductor may be penetrated by a rotated
magnetic field $\widetilde{B}$. Although a few works \cite{oldCS-B}
had previously addressed the problem of the interaction of an
external magnetic field with dense quark matter, none of these
studies considered the modification produced by the field on the gap
itself. However, as we have recently shown \cite{MCFL}, the gap
structure gets modified due to the penetrating field. To understand
this, notice that, although the condensate is
$\widetilde{Q}$-neutral, some of the quarks participating in the
pairing are $\widetilde{Q}$-charged and hence can couple to the
background field, which in turn affects the gap equations through
the Green functions of these $\widetilde{Q}$-charged quarks. Due to
the coupling of the charged quarks with the external field, the
color-flavor space is augmented by the $\widetilde{Q}$-charge
color-flavor operator, and consequently the CFL order parameter
splits in new independent pieces giving rise to a new phase, the
MCFL phase.

\section{The MCFL Gap Structure}\label{MCFL Gap}

The linear combination of the photon $A_{\mu}$ and a gluon
$G^{8}_{\mu}$ that behaves as a long-range field in the spin-zero
color superconductor is given by \cite{alf-raj-wil-99/537,
alf-berg-raj-NPB-02},

\begin{equation}
\widetilde{A}_{\mu}=\cos{\theta}A_{\mu}-\sin{\theta}G^{8}_{\mu},\label{1}
\end{equation}
while the orthogonal combination
$\widetilde{G}_{\mu}^8=\sin{\theta}A_{\mu}+\cos{\theta}G^{8}_{\mu}$
is massive. In the CFL phase the mixing angle $\theta$ is
sufficiently small ($\sin{\theta}\sim e/g\sim1/40$). Thus, the
penetrating field in the color superconductor is mostly formed by
the photon with only a small gluon admixture.

The unbroken $U(1)$ group corresponding to the long-range rotated
photon (i.e. the $\widetilde {U}(1)_{\rm e.m.}$) is generated, in
flavor-color space, by $\widetilde {Q} = Q \times 1 - 1 \times Q$,
where $Q$ is the electromagnetic charge generator. We use the
conventions $Q = -\lambda_8/\sqrt{3}$, where $\lambda_8$ is the 8th
Gell-Mann matrix. Thus our flavor-space ordering is $(s,d,u)$. In
the 9-dimensional flavor-color representation that we will use in
this paper (the color indexes we are using are (1,2,3)=(b,g,r)), the
$\widetilde{Q}$ charges of the different quarks, in units of
$\widetilde{e} = e \cos{\theta}$, are
\begin{equation}
\label{q-charges}
\begin{tabular}{|c|c|c|c|c|c|c|c|c|}
  \hline
  $s_{1}$ & $s_{2}$ & $s_{3}$ & $d_{1}$ & $d_{2}$ & $d_{3}$ & $u_{1}$ & $u_{2}$ & $u_{3}$ \\
  \hline
  0 & 0 & - & 0 & 0 & - & + & + & 0 \\
  \hline
\end{tabular}
\end{equation}

In the presence of an external rotated magnetic field the kinetic
part of the quarks' Lagrangian density must be rewritten using the
covariant derivative

\begin{equation}
L_{quarks}^{em}=\overline{\psi }(i\Pi_{\mu }\gamma ^{\mu })\psi \ ,
\label{quark-free-lag}
\end{equation}
with

\begin{equation}
\Pi _{\mu }=i\partial _{\mu
}+\widetilde{e}\widetilde{Q}\widetilde{A}_{\mu}  \ .
\label{pi-operator}
\end{equation}
where
\begin{equation}
\widetilde{Q}=\Omega _{+}-\Omega _{-}  \ .
 \label{Q-omeg-relat}
\end{equation}
is the rotated charge operator. The charge projectors

\begin{equation}
\Omega _{+}={\rm diag}(0,0,0,0,0,0,1,1,0) \ , \label{pos-omeg}
\end{equation}

\begin{equation}
\Omega _{-}={\rm diag}(0,0,1,0,0,1,0,0,0)  \ , \label{neg-omeg}
\end{equation}
and
\begin{equation}
\Omega _{0}={\rm diag}(1,1,0,1,1,0,0,0,1) \ ,
 \label{neut-omeg}
\end{equation}
obey the algebra

\begin{equation}
\Omega _{\eta }\Omega _{\eta ^{\prime }}=\delta _{\eta \eta ^{\prime
}}\Omega _{\eta },\qquad \eta ,\eta ^{\prime }=0,+,- \ .
\label{omeg-algeb}
\end{equation}

\begin{equation}
\Omega _{0}+\Omega _{+}+\Omega _{-}=1 \ . \label{omeg-sum}
\end{equation}

The rotated magnetic field naturally separates the quark fields
according to their $\tilde{Q}$ charge. The fermion field in the
$9\times9$ representation used above,
$\psi^{T}=(s_{1},s_{2},s_{3},d_{1},d_{2},d_{3},u_{1},u_{2},u_{3})$,
can then be written as the sum of three fields with zero, positive
and negative rotated  electromagnetic charges,

\begin{equation}
\psi =\psi _{(0)}+\psi _{(+)}+\psi _{(-)}  \ , \label{ferm-sum}
\end{equation}
where the $(0)$-, ($+/-)$-charged fields can be respectively written
in terms of the charge projectors as

\begin{equation}
\psi _{(0)}=\Omega _{0}\psi  \ ,\qquad \psi _{(+)}=\Omega _{+}\psi \
,\qquad \psi _{(-)}=\Omega _{-}\psi \ .  \label{fields-def}
\end{equation}

A strong magnetic field affects the flavor symmetries of QCD, as
different quark flavors have different electromagnetic charges. For
three light quark flavors, only the subgroup of $SU(3)_L \times
SU(3)_R$ that commutes with $Q$, the electromagnetic generator, is a
symmetry of the theory. Similarly, in the CFL phase a strong
$\widetilde{B}$ field should affect the symmetries of the theory, as
$\widetilde{Q}$ does not commute with the whole locked $SU(3)$
group. Based on the above considerations, and imposing that the
condensate should retain the highest degree of symmetry, we proposed
\cite{MCFL} the following ansatz for the gap structure in the
presence of a magnetic field

\begin{equation}
%\begin{eqnarray}
%\label{gapMCFL}
\Delta=\left(
\begin{array}{ccccccccc}
2\Delta^{'}_{S} & 0 & 0 & 0 & \Delta_{A}+\Delta_{S} & 0 & 0 & 0 & \Delta^{B}_{A}+\Delta^{B}_{S}\\
0 & 0 & 0 & \Delta_{S}-\Delta_{A} & 0 & 0 & 0 & 0& 0 \\
0 & 0 & 0 & 0 & 0 & 0 & \Delta^{B}_{S}-\Delta^{B}_{A} & 0 & 0 \\
0 & \Delta_{S}-\Delta_{A} & 0 & 0 & 0 & 0 & 0 & 0 & 0 \\
\Delta_{A}+\Delta_{S} & 0 & 0 & 0 & 2\Delta^{'}_{S} & 0 & 0 & 0 &  \Delta^{B}_{A}+\Delta^{B}_{S}\\
0 & 0 & 0 & 0 & 0 & 0& 0 & \Delta^{B}_{S}-\Delta^{B}_{A} & 0 \\
0 & 0 & \Delta^{B}_{S}-\Delta^{B}_{A} & 0 & 0 & 0 & 0 & 0 & 0 \\
0 & 0 & 0 & 0 & 0 & \Delta^{B}_{S}-\Delta^{B}_{A} & 0 & 0&0\\
\Delta^{B}_{A}+\Delta^{B}_{S} & 0 & 0 & 0 &
\Delta^{B}_{A}+\Delta^{B}_{S} & 0 & 0 &0&2\Delta^{''}_{S}
\label{order-parameter}
\end{array} \right)
\end{equation}

We call the reader's attention to the fact that despite the
$\widetilde{Q}$-neutrality of all the condensates, they can be
composed either by neutral or by charged quarks. Condensates formed
by $\widetilde{Q}$-charged quarks feel the field directly through
the minimal coupling of the background field $\widetilde{B}$ with
the quarks in the pair. A subset of the condensates formed by
$\widetilde{Q}$-neutral quarks, can feel the presence of the field
via tree-level vertices that couple them to charged quarks. The gaps
$\Delta^B_{A/S}$ are antisymmetric/symmetric combinations of
condensates composed by charged quarks and condensates formed by
this kind of neutral quarks. The gaps $\Delta_{A}$, as well as
$\Delta_{S}$, $\Delta^{'}_{S}$ and $\Delta^{''}_{S}$, on the other
hand, are antisymmetric and symmetric combinations of condensates
formed by neutral quarks that do not belong to the above subset. The
only way the field can affect them is through the system of highly
non-linear coupled gap equations. At zero field the CFL gap matrix
is recovered since in that case $\Delta^{B}_{A}=\Delta_{A}$ and
$\Delta^{B}_{S}=\Delta_{S}=\Delta^{'}_{S}=\Delta^{''}_{S}$.

Although the symmetry of the problem allows for four independent
symmetric gaps, the condensates $\Delta^{'}_{S}$ and
$\Delta^{''}_{S}$ are only due to subleading color symmetric
interactions, and as explained in the previous paragraph, they are
formed by neutral quarks that are not coupled to charged quarks, so
they belong to the same class as $\Delta_{S}$. Therefore, there is
no reason to expect that they will differ much from $\Delta_{S}$.
Hence, in a first approach to the problem, we will consider
$\Delta_S\simeq \Delta^{'}_{S} \simeq \Delta^{''}_{S}$.

 The order parameter (\ref{order-parameter}) implies the
following symmetry breaking pattern: $SU(3)_{\rm color} \times
SU(2)_L \times SU(2)_R \times U(1)_B \times U^{(-)}(1)_A \times
U(1)_{\rm e.m.} \rightarrow SU(2)_{{\rm color}+L+R} \times
{\widetilde {U}(1)}_{\rm e.m.}$. The $U^{(-)}(1)_A$ symmetry is
connected with the current which is an anomaly-free linear
combination of $s,d$ and $u$ axial currents
\cite{miransky-shovkovy-02}.  The locked $SU(2)$ corresponds to the
maximal unbroken symmetry, and as such it maximizes the condensation
energy. Notice that it commutes with the rotated electromagnetic
group ${\widetilde {U}(1)}_{\rm e.m.}$.

The counting of broken generators, after taking into account the
Anderson-Higgs mechanism, tells us that there are only five
Nambu-Goldstone bosons. One is associated to the breaking of the
baryon symmetry; three Goldstone bosons are associated to the
breaking of $SU(2)_A$, and another one associated to the breaking of
$U^{(-)}(1)_A$. All the Nambu-Goldstone bosons are
$\widetilde{Q}$-neutral. The number and properties of the lightest
particles in the MCFL have implications for the low-energy physics
of the phase. Since in her talk Cristina Manuel will address the
low-energy physics of the MCFL phase, I will not extend on this
topic in mine.

\section{Effective Action in the Presence of a Magnetic Field}

Let us construct the effective action of the system in the presence
of a magnetic field. With this aim, we will use a Nambu-Jona-Lasinio
(NJL) four-fermion interaction abstracted from one-gluon exchange
\cite{alf-raj-wil-99/537}. Although this simplified treatment
disregards the effect of the $\widetilde {B}$-field on the gluon
dynamics and assumes the same NJL couplings for both the situation
with and without magnetic field, it keeps the main attributes of the
theory, thereby providing the correct qualitative physics.

We start from the mean-field effective action
\begin{eqnarray}
I_{B}(\overline{\psi},\psi )
=\int\limits_{x,y}\{\frac{1}{2}[\overline{\psi }
_{(0)}(x)[G_{(0)0}^{+}]^{-1}(x,y)\psi _{(0)}(y)+\overline{ \psi
}_{(+)}(x)[G_{(+)0}^{+}]^{-1}(x,y)\psi _{(+)}(y) \nonumber \\
+\overline{\psi}_{(-)}(x)[G_{(-)0}^{+}]^{-1}(x,y)\psi _{(-)}(y)+
\overline{\psi } _{(0)C}(x)[G_{(0)0}^{-}]^{-1}(x,y)\psi
_{(0)C}(y)\nonumber \\
+\overline{ \psi }_{(+)C}(x)[G_{(+)0}^{-}]^{-1}(x,y)\psi _{(+)C}(y)
+\overline{\psi}_{(-)C}(x)[G_{(-)0}^{-}]^{-1}(x,y)\psi _{(-)C}(y)]
\nonumber \\
+\frac{1}{2}[\overline{\psi }_{(0)C}(x)\Delta ^{+}(x,y)\psi
_{(0)}(y)+h.c.] + \frac{1}{2}[\overline{\psi }_{(+)C}(x)\Delta
^{+}(x,y)\psi _{(-)}(y)\nonumber
\\
+\overline{\psi }_{(-)C}(x)\Delta ^{+}(x,y)\psi _{(+)}(y)+h.c.]\} \
, \label{b-coord-action}
\end{eqnarray}
where the external magnetic field has been explicitly introduced
through minimal coupling with the $\widetilde{Q}-$charged fermions.
The presence of the field is also taken into account in the diquark
condensate $\Delta^{+}=\gamma_{5}\Delta$, whose color-flavor
structure is given by Eq.(\ref{order-parameter}).

In (\ref{b-coord-action}) symbols in parentheses indicate neutral
$(0)$, positive $(+)$ or negative $(-)$ $\tilde{Q}-$charged quarks.
Supra-indexes $+$ or $-$ in the propagators indicate, as it is
customary, whether it is the inverse propagator of a field or
conjugated field respectively. Then, for example,
$[G_{(+)0}^{-}]^{-1}$ corresponds to the bare inverse propagator of
positively charged conjugate fields, and so on. The explicit
expressions of the inverse propagators are

\begin{equation}
\lbrack G_{(0)0}^{\pm}]^{-1}(x,y)=[i\gamma ^{\mu }\partial _{\mu
}-m\pm \mu \gamma ^{0}]\delta ^{4}(x-y) \ ,  \label{neut-x-inv-prop}
\end{equation}

\begin{equation}
\lbrack G_{(+)0}^{\pm }]^{-1}(x,y)=[i\gamma ^{\mu }\Pi ^{(+)}_{\mu
}-m\pm \mu \gamma ^{0}]\delta ^{4}(x-y) \ ,  \label{B-x-inv-prop+}
\end{equation}

\begin{equation}
\lbrack G_{(-)0}^{\pm }]^{-1}(x,y)=[i\gamma ^{\mu }\Pi ^{(-)}_{\mu
}-m\pm \mu \gamma ^{0}]\delta ^{4}(x-y) \ ,  \label{B-x-inv-prop-}
\end{equation}
with

\begin{equation}
\Pi ^{(\pm)}_{\mu }=i\partial _{\mu
}\pm\widetilde{e}\widetilde{A}_{\mu } \ . \label{piplusminus}
\end{equation}

Transforming the field-dependent quark propagators to momentum space
can be performed with the use of the Ritus' method, originally
developed for charged fermions \cite{Ritus:1978cj} and recently
extended to charged vector fields \cite{efi-ext}. In Ritus' approach
the diagonalization in momentum space of charged fermion Green's
functions in the presence of a background magnetic field is carried
out using the eigenfunction matrices $E_p(x)$. These are the wave
functions of the asymptotic states of charged fermions in a uniform
magnetic field and play the role in the magnetized medium of the
usual plane-wave (Fourier) functions $e^{i px}$ at zero field. Below
we introduce the basic properties of this transformation.

The transformation functions $E^{(\pm)}_{q}(x)$ for positively
($+$), and negatively ($-$) charged fermion fields are obtained as
the solutions of the field dependent eigenvalue equation

\begin{equation}
(\Pi^{(\pm)}\cdot\gamma)E^{(\pm)}_{q}(x)=E^{(\pm)}_{q}(x)(\gamma\cdot\overline{p}^{(\pm)})
\ , \label{eigenproblem}
\end{equation}
with $\overline{p}^{(\pm )}$ given by

\begin{equation}  \label{pbar+}
\overline{p}^{(\pm )}=(p_{0},0,\pm
\sqrt{2|\widetilde{e}\widetilde{B}|k},p_{3}) \ ,
\end{equation}
and
\begin{equation}
E^{(\pm)}_{q}(x)=\sum\limits_{\sigma }E^{(\pm)}_{q\sigma }(x)\delta
(\sigma ) \ , \label{9}
\end{equation}
with eigenfunctions

\begin{equation}
{E}^{(\pm)}_{p\sigma }(x)=\mathcal{N}%
_{n_{(\pm)}}e^{-i(p_{0}x^{0}+p_{2}x^{2}+p_{3}x^{3})}D_{n_{(\pm)}}(\varrho
_{(\pm)}) \ ,
 \label{Epsigma+}
\end{equation}
where $D_{n_{(\pm)}}(\varrho _{(\pm)})$ are the parabolic cylinder
functions with argument $\varrho _{(\pm)}$ defined by
\begin{equation}
\varrho _{(\pm)}=\sqrt{2|\widetilde{e}\widetilde{B}|}(x_{1}\pm p_{2}/\widetilde{e}%
\widetilde{B}) \ ,
 \label{rho+}
\end{equation}
and index $n_{(\pm)}$ given by
\begin{equation}  \label{normaliz-const+}
n_{(\pm)}\equiv n_{(\pm)}(k,\sigma)= k \pm \frac{\widetilde{e}\widetilde{B}}{2|%
\widetilde{e}\widetilde{B}|}\sigma-\frac{1}{2} \ , \qquad\qquad
n_{(\pm)}=0,1,2,...
\end{equation}
$k=0,1,2,3,...$ is the Landau level, and $\sigma$ is the spin
projection that can take values $\pm 1$ only. Notice that in the
lowest Landau level, $k=0$, only particles with one of the two spin
projections, namely, $\sigma=1$ for
positively charged particles, are allowed. The normalization constant $%
\mathcal{N}_{n_{(\pm)}}$ is
\begin{equation}  \label{normaliz-const}
\mathcal{N}_{n_{(\pm)}}=(4\pi| \widetilde{e}\widetilde{B}|)^{\frac{1}{4}}/%
\sqrt{n_{(\pm)}!} \ .
\end{equation}

In (\ref{9}) the spin matrices $\delta(\sigma )$ are defined as

\begin{equation}
\delta(\sigma )= {\rm diag}(\delta _{\sigma 1},\delta _{\sigma
-1},\delta _{\sigma 1},\delta _{\sigma -1}),\qquad \sigma =\pm 1 \ ,
\label{10}
\end{equation}
and satisfy the following relations

\begin{equation}
\delta \left( \pm \right) ^{\dagger }=\delta \left( \pm \right) \
,\qquad \delta \left( \pm \right) \delta \left( \pm \right) =\delta
\left( \pm \right)  \ ,\qquad \delta \left( \pm \right) \delta
\left( \mp \right) =0 \ ,
\end{equation}

\begin{equation}
\gamma ^{\shortparallel }\delta \left( \pm \right) =\delta \left(
\pm \right) \gamma ^{\shortparallel },\quad \gamma ^{\bot }\delta
\left( \pm \right) =\delta \left( \mp \right) \gamma ^{\bot } \ .
\label{24}
\end{equation}
In Eq. (\ref{24}) the notation $\gamma ^{\shortparallel }=(\gamma
^{0},\gamma ^{3})$ and $\gamma ^{\bot }=(\gamma ^{1},\gamma ^{2})$
was used.

 The functions $E^{(\pm)}_{p}$ are complete
\begin{equation}  \label{complete-Ep}
\sum_{k}\int dp_{0}dp_{2}dp_{3}{E}^{(\pm)}_{p}(x){\overline{E}}%
^{(\pm)}_{p}(y)=(2\pi)^{4}\delta^{(4)}(x-y) \ ,
\end{equation}
and orthonormal,
\begin{equation}  \label{ortho-Ep}
\int_{x}{\overline{E}}^{(\pm)}_{p^{\prime}}(x){E}^{(\pm)}_{p}(x)=(2\pi)^{4}%
\Lambda_{k}\delta_{kk^{\prime}}\delta(p_{0}-p^{\prime}_{0})
\delta(p_{2}-p^{\prime}_{2})\delta(p_{3}-p^{\prime}_{3})
\end{equation}
with the $(4\times4)$ matrix $\Lambda_{k}$ given by

\begin{eqnarray}
\Lambda_{k}= \left\{
\begin{array}{cc}
\delta(\sigma= {\rm sgn}[eB]) \qquad\qquad for \qquad  k=0,\\
\qquad I \qquad\qquad\qquad\qquad for \qquad k>0 .
\end{array} \right.
\end{eqnarray}

The matrix structure $\Lambda_{k}$ was recently introduced in Ref.
\cite{Leung05}. It had been previously omitted in the orthonormal
condition of the $E_p(x)$ functions given in Refs.
\cite{Ritus:1978cj, efi-ext, orthonormality}. Nevertheless, it
should be underlined that this matrix only appears in the zero
Landau level contribution, and consequently it enters as an
irrelevant multiplicative factor in the Schwinger-Dyson equations in
the lowest Landau level approximation. Thus, all the results
obtained in the works \cite{Ritus:1978cj, efi-ext, orthonormality}
remain valid. In Eqs. (\ref{complete-Ep})-(\ref{ortho-Ep}) we
introduced the notation $\overline{E}_{p}^{(\pm)}(x)=\gamma
_{0}({E}_{p}^{(\pm)}(x))^{\dag }\gamma _{0}$.

Under the $E_p(x)$ functions, positively ($\psi _{(+)}$), negatively
($\psi _{(-)}$) charged fields transform according to
\begin{equation}
\psi _{(\pm)}(x)=\sum_{k}\int dp_{0}dp_{2}dp_{3}E_{p}^{(\pm)}(x)\psi
_{(\pm)}(p) \ ,\label{psi+-transf}
\end{equation}
\begin{equation}
\overline{\psi }_{(\pm)}(x)=\sum_{k}\int dp_{0}dp_{2}dp_{3}\overline{\psi }%
_{(\pm)}(p)\overline{E}_{p}^{(\pm)}(x)  \ . \label{psibar+-transf}
\end{equation}

One can show that
\begin{equation}  \label{gamma-piplus-propert}
[\gamma_{\mu}(\Pi_{(+)\mu}\pm \mu\delta_{\mu0})- m]{E}^{(+)}_{p}(x)={E}%
^{(+)}_{p}(x)[\gamma_{\mu}(\overline{p}^{(+)}_{\mu}\pm
\mu\delta_{\mu0})- m] \ ,
\end{equation}
and
\begin{equation}  \label{gamma-piminus-propert}
[\gamma_{\mu}(\Pi_{(-)\mu}\pm \mu\delta_{\mu0})- m]{E}^{(-)}_{p}(x)={E}%
^{(-)}_{p}(x)[\gamma_{\mu}(\overline{p}^{(-)}_{\mu}\pm
\mu\delta_{\mu0})- m] \ .
\end{equation}

The conjugate fields transform according to,

\begin{equation}  \label{conjpsiplustransf}
\psi_{(+)C}(x)=\sum_{k}\int
dp_{0}dp_{2}dp_{3}E^{(-)}_{p}(x)\psi_{(+)C}(p),
\end{equation}
\begin{equation}  \label{conjpsiminustransf}
\psi_{(-)C}(x)=\sum_{k}\int
dp_{0}dp_{2}dp_{3}E^{(+)}_{p}(x)\psi_{(-)C}(p) \ .
\end{equation}

After transforming to momentum space one can introduce Nambu-Gorkov
fermion fields of different $\tilde{Q}$ charges. They are the
$\tilde{Q}$-neutral Gorkov field

\begin{equation}
\Psi _{(0)}=\left(
\begin{array}{c}
\psi _{(0)} \\
\psi _{(0)C}
\end{array}
\right) \ ,
\end{equation}
the positive
\begin{equation}
\Psi _{(+)}=\left(
\begin{array}{c}
\psi _{(+)} \\
\psi _{(-)C}
\end{array}
\right) \ ,
\end{equation}
and the negative one
\begin{equation}
\Psi _{(-)}=\left(
\begin{array}{c}
\psi _{(-)} \\
\psi _{(+)C}
\end{array}
\right) \ .
\end{equation}

Using them, the Nambu-Gorkov effective action in the presence of a
constant magnetic field $\widetilde{B}$ can be written as

\begin{eqnarray}  \label{b-action}
I^{B}(\overline{\psi},\psi)
=\frac{1}{2}\int\frac{d^{4}p}{(2\pi)^{4}} \overline{\Psi}%
_{(0)}(p){\cal S}^{-1}_{(0)}(p)\Psi_{(0)}(p)\nonumber\\
+\frac{1}{2}\int\frac{d^{4}p}{(2\pi)^{4}}
\overline{\Psi}_{(+)}(p){\cal S}^{-1}_{(+)}(p)\Psi_{(+)}(p)+\frac{1}{2}\int\frac{d^{4}p}{%
(2\pi)^{4}} \overline{\Psi}_{(-)}(p){\cal
S}^{-1}_{(-)}(p)\Psi_{(-)}(p) \ ,
\end{eqnarray}
where
\begin{eqnarray}  \label{p-neutr-inv-propg}
{\cal S}^{-1}_{(0)}(p)=\left(
\begin{array}{cc}
[G_{(0)0}^{+}]^{-1}(p) & \Delta_{(0)}^{-} \\
&  \\
\Delta_{(0)}^{+} & [G_{(0)0}^{-}]^{-1}(p)
\end{array}
\right) \ ,
\end{eqnarray}

\begin{eqnarray}  \label{p-posit-inv-propg}
{\cal S}^{-1}_{(+)}(p)=\left(
\begin{array}{cc}
[G_{(+)0}^{+}]^{-1}(p) & \Delta_{(+)}^{-} \\
&  \\
\Delta_{(+)}^{+} & [G_{(+)0}^{-}]^{-1}(p)
\end{array}
\right) \ ,
\end{eqnarray}
\begin{eqnarray}  \label{p-negat-inv-propg}
{\cal S}^{-1}_{(-)}(p)=\left(
\begin{array}{cc}
[G_{(-)0}^{+}]^{-1}(p) & \Delta_{(-)}^{-} \\
&  \\
\Delta_{(-)}^{+} & [G_{(-)0}^{-}]^{-1}(p)
\end{array}
\right) \ ,
\end{eqnarray}
with

\begin{equation}
\Delta_{(+)}^{+}= \Omega_{-}\Delta^{+}\Omega_{+},
\end{equation}

\begin{equation}
\Delta_{(-)}^{+}= \Omega_{+}\Delta^{+}\Omega_{-},
\end{equation}

\begin{equation}
\Delta_{(0)}^{+}= \Omega_{0}\Delta^{+}\Omega_{0},
\end{equation}

Notice that to form the positive (negative) Nambu-Gorkov field we
used the positive (negative) fermion field and the charge conjugate
of the negative (positive) field. This is done so that the rotated
charge of the up and down components in a given Nambu-Gorkov field
be the same. This way to form the Nambu-Gorkov fields is mandated by
what kind of field enters in a given condensate term, which in turn
is related to the neutrality of the fermion condensate $\langle
\overline{\psi}_{C}\psi\rangle$ with respect to the rotated
$\tilde{Q}$-charge.

In momentum space the bare inverse propagator for the neutral field
is

\begin{equation}
\lbrack G_{(0)0}^{\pm }]^{-1}(p)=[\gamma _{\mu }(p_{\mu }\pm \mu
\delta _{\mu 0})-m] \ , \label{neut-bareprop+-}
\end{equation}
where the momentum is the usual $p=(p_{0},p_{1},p_{2},p_{3})$ of the
case with no background field.

For positively and negatively charged fields the bare inverse
propagators are
\begin{equation}
\lbrack G_{(+)0}^{\pm }]^{-1}(p)=[\gamma _{\mu }(\overline{p}%
_{\mu }^{(+)}\pm \mu \delta _{\mu 0})-m] \ , \label{pos-bareprop}
\end{equation}
and
\begin{equation}
\lbrack G_{(-)0}^{\pm }]^{-1}(p)=[\gamma _{\mu }(\overline{p}%
_{\mu }^{(-)}\pm \mu \delta _{\mu 0})-m]  \ \label{neg-bareprop}
\end{equation}
respectively.
\section{Gap Solutions}

The main question we would like to address now is: Can we find a
region of magnetic fields where the gaps $\Delta_{A}$ and
 $\Delta^B_{A}$, (or $\Delta_{S}$ and
 $\Delta^B_{S}$), differ enough from each other that the system is
 not in the CFL phase anymore, but in the MCFL phase? To explore the possible answer to this
 question we need to solve the gap equations derived from the
 Nambu-Gorkov effective action (\ref{b-action}).

In coordinate space the  QCD gap equation reads

\begin{equation}
\label{gap-eq} \Delta^+(x,y) = i\frac{g^2}{4} \lambda_A^T\,
\gamma^\mu\, S_{21}(x,y) \gamma^\nu\, \lambda_B D^{AB}_{\mu \nu}
(x,y) \ ,
\end{equation}
where $S_{21}(x,y)$ is the off-diagonal part of the Nambu-Gorkov
fermion propagator in coordinate space and, for simplicity, we have
omitted explicit color and flavor indices in the gap and fermion
propagator. Here $D^{AB }_{\mu \nu}$ is the gluon propagator.

In a NJL model the gap equation can be obtained from
Eq.~(\ref{gap-eq}) simply by substituting the gluon propagator by

\begin{equation}D^{AB}_{\mu \nu} (x,y) = \frac{1}{\Lambda^2} \,g_{\mu \nu}\,
\delta^{AB} \,\delta^{(4)}(x -y) \ .\end{equation}

The NJL model is characterized by a coupling constant $g$ and an
ultraviolet cutoff $\Lambda$. The ultraviolet cutoff should be much
larger than any of the energy scales of the system, typically the
chemical potential. In the presence of a magnetic field we should
also assume that $\Lambda$ is larger than the magnetic energy
$\sqrt{{\tilde e}{\tilde B} }$. In other studies of color
superconductivity within the NJL model, the values of $g$ and
$\Lambda$ are chosen to match some QCD vacuum properties, thus
hoping to get in such a way correct approximated quantitative
results of the gaps. We will follow the same philosophy here,
noticing however that this  completely ignores the effect of the
magnetic field on the gluon dynamics.

To solve the gap equation (\ref{gap-eq}) for the whole range of
magnetic-field strengths we need to use numerical methods. We have
found, however, a situation where an analytical solution is
possible. This corresponds to the case $\widetilde{e}\widetilde{B}
\gtrsim \mu^2/2$. Taking into account that the leading contribution
to the gap solution comes from quark energies near the Fermi level,
it follows that for fields in this range only the LLL ($l=0$)
contributes.

Using the approximation $\Delta^B_A \gg \Delta^B_S, \Delta_A$, and
$\Delta_A \gg \Delta_S$, the gap equations decouple and the equation
for $\Delta^B_A$ is
\begin{eqnarray}
\label{maingeq} \Delta^B_A & \approx & \frac{g^2}{3 \Lambda^2}
\int_{\Lambda} \frac{d^3 q}{(2 \pi)^3} \frac{
\Delta^B_A}{\sqrt{(q-\mu)^2 + 2 (\Delta^B_A)^2 }} \nonumber
 \\
& + & \frac{g^2 \widetilde{e}\widetilde{B}}{3 \Lambda^2}
\int_{-\Lambda}^{\Lambda} \frac{d q}{(2 \pi)^2} \frac{
\Delta^B_A}{\sqrt{(q-\mu)^2 + (\Delta^B_A)^2 }} ,
\end{eqnarray}
where the first/second term in the r.h.s. of Eq.(\ref{maingeq})
corresponds to the contribution of $\widetilde{Q}$-neutral/charged
quark propagators, respectively. For the last one, we dropped all
Landau levels but the lowest, as we are interested in the leading
term.

The solution of Eq.~(\ref{maingeq}) reads
\begin{equation}
\label{gapBA} \Delta^B_A \sim 2 \sqrt{\delta \mu} \, \exp{\Big( -
\frac{3 \Lambda^2 \pi^2} {g^2 \left(\mu^2 + \widetilde{e}
\widetilde{B} \right)} \Big) } \ ,
\end{equation}
with $\delta \equiv \Lambda - \mu$. It can be compared with the
antisymmetric CFL gap \cite{review}
\begin{equation}
\label{gapCFL} \Delta^{\rm CFL}_A \sim 2 \sqrt{\delta \mu} \,
\exp{\Big( -\frac{3 \Lambda^2 \pi^2} {2 g^2 \mu^2} \Big) } \ .
\end{equation}

In this approximation the remaining gap equations read
\begin{eqnarray}
\label{symgeq} \Delta^B_S & \approx & -\frac{g^2}{6 \Lambda^2}
\int_{\Lambda} \frac{d^3 q}{(2 \pi)^3}
\frac{ \Delta^B_A}{\sqrt{(q-\mu)^2 + 2 (\Delta^B_A)^2 }} \nonumber\\
& + & \frac{g^2 \widetilde{e}\widetilde{B}}{6 \Lambda^2}
\int_{-\Lambda}^{\Lambda} \frac{d q}{(2 \pi)^2} \frac{
\Delta^B_A}{\sqrt{(q-\mu)^2 + (\Delta^B_A)^2 }}   \ ,
\end{eqnarray}

\begin{eqnarray}
\label{antisymme-gapeqapp}
 \Delta_A & \approx & \frac {g^2}{4 \Lambda^2}  \int_{\Lambda} \frac{d^3q}{(2 \pi)^3}
 \Big( \frac {17}{9} \frac{\Delta_A}{ \sqrt{ (q - \mu)^2 +\Delta_A^2 }}
\nonumber \\
& + &  \frac{7}{9} \frac{\Delta_A}{ \sqrt{(q-\mu)^2 + 2
(\Delta^B_A)^2 }  }
 \Big) \ ,
 \end{eqnarray}
and

\begin{eqnarray}
\label{symme-gapeqapp}
 \Delta_S & \approx &  \frac {g^2}{18 \Lambda^2}  \int_{\Lambda} \frac{d^3q}{(2 \pi)^3}
\Big(  \frac{\Delta_A}{ \sqrt{ (q - \mu)^2 +\Delta_A^2 }} \nonumber
\\
& - &   \frac{\Delta_A}{ \sqrt{(q-\mu)^2 + 2 (\Delta^B_A)^2 } }
\Big) \ .
 \end{eqnarray}

We express below the solution of these gap equations as ratios over
the CFL antisymmetric and symmetric gaps
\begin{equation}
 \frac{\Delta_A}{\Delta^{\rm CFL}_A}  \sim \frac{1}{2^{(7/34)}} \exp{\Big(-\frac{36}{17 x}  + \frac{21}{17}\frac{1}{x (1 + y)} + \frac{3}{2x}
\Big) } \ ,
\end{equation}
where $x \equiv g^2 \mu^2/\Lambda^2 \pi^2$, and $y \equiv
\widetilde{e}\widetilde{B}/\mu^2$, and
\begin{eqnarray}
\frac{\Delta^B_S}{\Delta^{\rm CFL}_S} & \sim &
\frac{\Delta^B_A}{\Delta^{\rm CFL}_A} \left(
\frac{3}{4} + \frac{9}{2 x \ln{2}} \frac{y -1}{y+1} \right) \ ,\\
\frac{\Delta_S}{\Delta^{\rm CFL}_S} & \sim &
\frac{\Delta_A}{\Delta^{\rm CFL}_A} \frac 32 \left(1 - \frac{4}{1+y}
\right) \ .
\end{eqnarray}

Note that our analytic solutions are only valid at strong magnetic
fields. The lower value $\widetilde{e}\widetilde{B} \sim \mu^2/2$
corresponds to $\widetilde{e}\widetilde{B} \sim (0.8 - 1.1) \cdot
10^{18}$G, for $\mu \sim 350- 400$ MeV. For fields of this order and
larger the $\Delta^B_A$ gap is larger than $\Delta^{\rm CFL}_A$ at
the same density values. Nevertheless, we have estimated (see the
details in Cristina Manuel's talk in the proceedings) that the
separation between CFL and MCFL will take place already at fields
$\sim$ $10^{16}G$.

All the gaps feel the presence of the external magnetic field. The
effect of the magnetic field in $\Delta^{B}_{A}$ is to increase the
density of states, which enters in the argument of the exponential
as typical of a BCS solution. The density of states appearing in
(\ref{gapBA}) is just the sum of those of neutral and charged
particles participating in the given gap equation (for each Landau
level, the density of states around the Fermi surface for a charged
quark is $\widetilde{e}\widetilde{B}/2 \pi^2$).

All the $\widetilde{Q}$-charged quarks have common gap
$\Delta^{B}_{A}$. Hence, the densities of the charged quarks are all
equal. As two of these quarks have positive $\widetilde{Q}$ charge,
while the other two have it negative, the $\widetilde{Q}$ neutrality
of the medium is guaranteed without having to introduce any electron
density.

\section{Conclusions}

In this paper, we have shown that a magnetic field leads to the
formation of a new color-flavor locking phase, characterized by a
smaller vector symmetry than the CFL phase. The essential role of
the penetrating magnetic field is to modify the density of states of
charged quarks on the Fermi surface. To better understand the
relevance of this new phase in astrophysics we need to explore the
region of moderately strong magnetic fields
$\widetilde{e}\widetilde{B}< \mu^2/2$, which requires to carry out a
numerical study of the gap equations including the effect of higher
Landau levels. Because the total density of states around the Fermi
surface for charged particles does not vary monotonically with the
number of Landau levels, we still expect to find a meaningful
splitting of the gaps at these fields and therefore a qualitative
separation between the CFL and MCFL phases.

\textbf{Acknowledgments}\\
The work of E.J.F. and V.I. was supported in part by NSF grant
PHY-0070986, and C.M. was supported by MEC under grant
FPA2004-00996.


\begin{thebibliography}{10}

\bibitem{alf-raj-wil-99/537}
M. Alford , K. Rajagopal  and F. Wilczek, \emph{"Color Flavor
Locking and Chiral Symmetry Breaking ,"} \emph{Nucl. Phys. B}
\textbf{537}, 443 (1999).

\bibitem{magnetars}
C.~Thompson and R.~C.~Duncan, \emph{``The Soft Gamma Repeaters as
Very Strongly Magnetized Neutron Stars. 2. Quiescent Neutrino,
X-ray, and Alfven Wave Emission,''} \emph{Astrophys. J.}  {\bf 473},
322 (1996).

\bibitem{lugones/0504454}German Lugones, \emph{"Magnetic Fields in High-Density Stellar Matter,
"} [{\tt astro-ph/0504454}].

\bibitem{virial}
L. Dong and S.L. Shapiro, \emph{"Cold equation of state in a strong
magnetic field - Effects of inverse beta-decay,"} \emph{Astrophys.
J.} {\bf 383}, 745 (1991).

\bibitem{MCFL}
 E.~J.  Ferrer ,V. de la Incera  and C. Manuel,
 \emph{"Magnetic Color Flavor Locking Phase in High Density QCD,''}
\emph{PRL} {\bf 95}, 152002 (2005).


%\bibitem{casal-gat-plb-464}R. Casalbuoni and R. Gatto, \emph{"Effective
%theory for color-flavor locking in high density QCD," }  \emph{Phys.
%Lett. B} {\bf 464}, 111 (1999).

\bibitem{oldCS-B} E. V. Gorbar,\emph{"Color superconductivity in an external magnetic field,"}
\emph{Phys. Rev. D} {\bf 62}, 014007 (2000); K. Iida and G. Baym,
\emph{"Superfluid phases of quark matter. III. Supercurrents and
vortices,"} \emph{Phys. Rev. D} {\bf66}, 014015 (2002); I. Giannakis
and H-C Ren,\emph{"The Ginzburg-Landau theory and the surface energy
of a colour superconductor,"} \emph{Nucl. Phys. B}  {\bf 669}, 462
(2003).

\bibitem{alf-berg-raj-NPB-02}
Alford M,  Berges J, and Rajagopal K, \emph{"Magnetic fields within
color superconducting neutron star cores,"}  \emph{Nucl. Phys. B}
\textbf{571}, 269 (2000).

\bibitem{miransky-shovkovy-02}V.~A.~Miransky, and I.~A.~Shovkovy,
\emph{"Magnetic catalysis and anisotropic confinement in QCD,"}
\emph{Phys. Rev. D} {\bf 66}, 045006 (2002).

\bibitem{Ritus:1978cj}
V.I. Ritus,\emph{"Radiative Corrections in Quantum Elctrodynamics
with Intense Field and their Analytical Properties,"} \emph{
Ann.Phys.} {\bf69}, 555 (1972).

\bibitem{efi-ext}
E. Elizalde, E. J. Ferrer, and V. de la Incera, \emph{"Neutrino
Self-Energy and Index of Refraction in Strong Magnetic Field: A New
Approach,"} \emph{Ann. of Phys.} {\bf 295}, 33 (2002);
\emph{"Neutrino Propagation in a Strongly Magnetized Medium,"}
\emph{Phys. Rev. D} {\bf 70}, 043012 (2004).

\bibitem{Leung05}
C. N. Leung, and S.-Y. Wang, \emph{"Gauge independent approach to
chiral symmetry breaking in a strong magnetic field,"}
hep-ph/0510066.

\bibitem{orthonormality}
D.-S Lee, C. N. Leung and Y. J. Ng, \emph{"Chiral symmetry breaking
in a uniform external magnetic field,"} \emph{Phys. Rev. D} {\bf
55}, 6504 (1997); E. J. Ferrer, and V. de la Incera,
\emph{"Ward-Takahashi Identity with External Field in Ladder QED,"}
\emph{Phys. Rev. D} {\bf 58}, 065008 (1998); \emph{"Magnetic
Catalysis in the Presence of Scalar Fields} \emph{Phys. Lett. B}
{\bf 481}, 287 (2000); E. Elizalde, E. J. Ferrer, and V. de la
Incera, \emph{"Beyond-Constant-Mass-Approximation Magnetic Catalysis
in the Gauge Higgs-Yukawa Model,"} \emph{Phys. Rev. D} {\bf 68},
096004 (2003).

\bibitem{review}
  K.~Rajagopal and F.~Wilczek,\emph{"The condensed matter physics of
  QCD,"} hep-ph/0011333.
\end{thebibliography}
\end{document}